\documentclass[12pt]{article}

\def\rdots{\mathinner{\mkern1mu\raise1pt\vbox{\kern1pt\hbox{.}}\mkern2mu
   \raise4pt\hbox{.}\mkern2mu\raise7pt\hbox{.}\mkern1mu}}
\newcommand{\Z}{{\rm Z\kern-.35em Z}}
\newcommand{\bP}{{\rm I\kern-.15em P}}
\newcommand{\Q}{\kern.3em\rule{.07em}{.65em}\kern-.3em{\rm Q}}
\newcommand{\R}{{\rm I\kern-.15em R}}
\newcommand{\h}{{\rm I\kern-.15em H}}
\newcommand{\C}{\kern.3em\rule{.07em}{.65em}\kern-.3em{\rm C}}
\newcommand{\T}{{\rm T\kern-.35em T}}

\newcommand{\be}{\begin{equation}}
\newcommand{\ee}{\end{equation}}

\newcommand{\ve}{\varepsilon}

\newcommand{\ga}{\gamma}

\newcommand{\ra}{\rightarrow}

\newcommand{\de}{\delta}
\newcommand{\La}{\Lambda}

\newcommand{\al}{\alpha}

\begin{document}

\openup 1.5\jot
\centerline{For the Quantum Heisenberg Ferromagnet, }
\centerline{Tao to the Proof of a Phase Transition}

\vspace{2in}
\centerline{Paul Federbush}
\centerline{Department of Mathematics}
\centerline{University of Michigan}
\centerline{Ann Arbor, MI 48109-1109}
\centerline{(pfed@umich.edu)}

\vfill\eject

\centerline{\underline{Abstract}}
\bigskip

We present the outline of a proof for the 3-d phase transition, which we hope to carry forth.  At the same time this paper provides some physical understanding of the phase transition, in the flavor of relatively simple arguments from an undergraduate statistical mechanics course.  A number of directions for mathematical research, interesting in their own right, will be suggested by aspects of the development.  We hope and believe that readers will be enticed by the {\it naturalness} and {\it beauty} of the path; some perhaps even, big game veterans, sniffing the quarry, will be ready to join the hunt.

The central construct views the trace, $Tr(e^{-\beta H})$, as a lattice gas of polymers, each representing a cycle in the permutation group, with hard core interactions.  The activities of the polymers have expressions as arising from the main conjecture of the paper.  The estimates lead to a phase-transition in 3-d, but not in 2-d.  This occurs via the same argument that a random walk in 2-d has certain return to the origin, but not so for a random walk in 3-d.

\vfill\eject

\section{Introduction.}

\ \ \ \ \ \ \ This work does not depend on our previous poking at the Quantum Heisenberg system, [1], [2].  We only learned from this previous study the surprisingly relevant relation between solutions of the heat equation and quantities in this model.  There is the precise rigorous relation of Eq. (19) of [1]; and the numerical approximations of [1], which [2] makes feeble effort to justify.  The key conjecture of this paper, as given in Section 3, is so inspired.

We work in $d$ dimensions, on a cubical periodic lattice, $\La$, of side $L$, so the total number of lattice sites is $N = L^d$.  The Hilbert space, $\cal H$, splits into sectors ${\cal H}_i$, $i=0,...,N$, where in ${\cal H}_i$ there are $i$ spins up.  The Hamiltonian, $H$, is given as
\be	H= -\sum_{i \sim j} (I_{ij} - 1)		\ee
$I_{ij}$ interchanges the spins of two neighboring sites $i$ and $j$ of the lattice $\La$.  We will sometimes view $H$ as an operator on $\cal H$, and sometimes as an element of the group algebra of the permutation group on the $N$ vertices of $\La$, allowing $I_{ij}$ to interchange vertices $i$ and $j$.

In what follows most of the development is not precise and rigorous, hand waving in nature.  The conjectures are not precise either.  We are far from a mathematically rigorous treatment.  However, the independence of the arguments on precise details also means that a mathematically honest proof of the phase transition along these lines will not depend on obtaining proofs of the conjectures in a very circumscribed form.  I.e., estimates of the flavor of our conjectures should work.

\section{Strategy for Proving a Phase Transition.}
\setcounter{equation}{0}

\ \ \ \ \ \ In this Section all arguments are precise, and results proven or easy to prove.  $H$ is taken as an operator on $\cal H$.  We let $Tr(e^{-\beta H})_{L,i}$ be the trace of $e^{-\beta H}$ restricted to ${\cal H}_i$, with $L$ the edge size (which we will vary).  Let 
\be F_\beta(L, n) = \sum^n_{i=0} \ Tr(e^{-\beta H})_{L,i}  . \ee
We note
\be	F_\beta(L, N) = Tr(e^{-\beta H}).	\ee

\bigskip
\noindent
\underline{Theorem.}  {\it Let $r < \frac 1 2$ and $\beta$ be fixed.  Then if}
\be	\frac {F_\beta(L, [rN])} {F_\beta(L,N)} > \ga > 0	\ee
{\it for some $\ga$ and all large enough $L$, there is spontaneous magnetization for such $\beta$.  Here [s] is the largest integer in $s$, and $N=L^d$.}
\bigskip
\bigskip

We have not used a standard definition of spontaneous magnetization.  We consider the two-point correlation function:
\be 	\rho_L(i, j) = \frac{Tr(e^{-\beta H}\sigma_{iz}\sigma_{jz})_L} {Tr(e^{-\beta H})_L} \ee
where the subscript, $L$, of course, indicates the edge size of $\La$.  We argue for spontaneous magnetization by excluding the existence of a $d(|i - j|)$ with
\be	\lim_{x \ra \infty} d(x) = 0	\ee
for which
\be	|\rho_L(i,j)| < d (|i - j|)    \ee
if
\be	|i-j| < \frac L 2 \ .	\ee
That is, we will show Eq. (2.3) implies there is no $d(x)$ satisfying Eq. (2.5) - (2.7).  We take this as the definition of spontaneous magnetization.  The limitation Eq. (2.7) is due to working in a periodic domain.

We assume, by contradiction, the existence of such a $d$ in the presence of Eq. (2.3) being satisfied.  We let $H' = \displaystyle{\sum_{i\in\La}} \; \sigma_{zi}$ and consider
\be	A_L(\delta) = \frac{Tr(e^{-\beta H - \delta H'})_L} {Tr(e^{-\beta H})_L} \ .\ee
One easily deduces the chain of inequalities 
\be	1 + t\delta^2 \sum_{i,j \in \La} d(|i-j|) > A_L(\delta) > 1+q  \de^2 N^2_L		\ee
with $t, q > 0$, and in the limit $\de$ goes to zero.  The right inequality in (2.9) comes from (2.3) and (2.8).  The left inequality in (2.9) comes from (2.8) and the definition of $\rho_L(i, j)$, expanding the exponent in $H'$ to second order.  The inequalities are inconsistent from
\be	\sum_{i,j \in \La} d(|i-j|) \le  a(\ve)N + \ve N^2	\ee
for each $\ve > 0$.

\bigskip
\bigskip

\section{The Central Approximation}
\setcounter{equation}{0}

\ \ \ \ \ \ In this section we view $H$ as an element of the group algebra of the permutation group on the $N$ vertices of $\La$.  Then
\be e^{-\beta H} = \sum_\al \tilde C_\al G_\al	\ee
where $G_\al$ is an element of the permutation group on $N$ letters.  We also want the lattice Laplacian heat equation Greens function
\be	g_\beta(i,j) = (e^{\beta \Delta})_{ij}		\ee
where, naturally, periodic boundary conditions are imposed in our periodic lattice.  We let $G_\al$ map the vertices as
\be	G_\al : i \ra i_\al, \ \ \ \ \ i=1,...,N	\ee
The approximation we now conjecture is
\be	\tilde C_\al \begin{array}[t]{c}  
{\displaystyle\longrightarrow}\\
{\scriptstyle{\beta \ra \infty}}
\end{array} \tilde C \prod^N_{i=1} g_\beta (i, i_\al) \ .  \ee
But we are not going to be precise in what sense the right side approximates the left side (the type of convergence).  We will in fact replace the left side by the right side in expressions we use from now on.  What we desire in a rigorous form of (3.4) is a result that enables the remaining proof to proceed.  We have ideas how to mathematically prove approximations similar to (3.4) and plan to work on them as the first step in rigorizing the present paper.

\bigskip
\bigskip
\section{The Polymers}
\setcounter{equation}{0}

\ \ \ \ \ \ Each permutation group element, $G$, has associated to it in a 1-1 way a partition of the $|\La| = N$ vertices, within each subset of the partition being given a specific cyclic ordering.  If $S$ is a subset of the partition with $k$ vertices, then $S$ may be given as
\be	\{i, Gi, G^2  i, ..., G^{k-1} i \} = S	\ee
for any $i$ in $S$.  $S$, of course, corresponds to a $k$-cycle of $G$.  We label the vertices in $S$ by $\al_1, \al_2, ..., \al_k$ with $G\al_i = \al_{i+1},  \ \ i < k$, $G\al_k = \al_1$.  To this $k$-cycle $S$ it is natural from (3.4) to associate an ``{\it activity}" $e_S$ by
\be
e_S = \left( \prod^{k-1}_{i=1} \ g_\beta (\al_i, \al_{i+1})\right) g_\beta(\al_1, \al_k).	\ee
We have constructed a ``{\it polymer}" with vertices, $\{\al_i\}$, and activity, $e_S$.  This we call a ``$k$-polymer".

We now consider the sum over all possible $k$-polymers through vertex $i$, each times its activity.  This leads to a sum 
\be
\sum_{\ga_2,\ga_3,...,\ga_k} g_\beta(i, \ga_2) g_\beta( \ga_2, \ga_3) \cdots \ g_\beta( \ga_{k-1}, \ga_k) g_\beta(i,\ga_k)
\ee
where the vertices $i, \ga_2, \dots, \ga_k$ are restricted to be distinct.  We estimate the sum in (4.3) to be
\be	\sim \ \ \frac 1{(\sqrt{\beta})^d} \cdot \frac 1{(\sqrt{k})^d} .	\ee
We argue this by viewing the sum in (4.3) to be a random walk in $d$ dimensions, with $k$ steps, and step size $\sim  \sqrt{\beta}$.  The random walk will then have travelled a mean-square distance $\sim \ \sqrt{\beta}\  \sqrt{k}$ and thus in $d$-dimensions have probability as given in (4.4) to return to origin, $i$.

\bigskip
\bigskip
\section{Statistical Mechanics of the Trace}
\setcounter{equation}{0}

\ \ \ \ \ \ For any element, $G$, of the permutation group we define $m(G)$ to be the number of cycles in $G$.  Referring to equation (3.1) we find the expression for the trace, Tr$(e^{-\beta H})$
\be	{\rm Tr}(e^{-\beta H}) = \sum_\al \ 2^{m(G_\al)} \tilde C_\al .	\ee
The factor of 2 associated to each cycle is from the choice of spin up or down.  Each vertex in a given cycle must have same value of spin.  (We evaluate the trace in the product of spin up, spin down bases, as usual.)  

We let $G_\al$ contain $s_\al(n)$ $n$-cycles.  Thus one must have
\be	\sum_{n=1} s_\al(n) \cdot n = L^d \ .		\ee
We play a usual statistical mechanics game of approximating the sum in (5.1) by keeping only terms with fixed values of the $s(n)$; and then maximizing this subsum of (5.1) over choices of the $s(n)$.  For a given $s(n)$ value we sum over the $s(n)$ choices of $n$-cycles using Boltzmann statistics
\be
\frac{(L^d)^{s(n)}}{s(n)!}  \cdot \left( \frac 1{(\beta n)^{d/2}} \right)^{s(n)} \cdot \frac 1{n^{s(n)}} \ .	\ee
The factorial arises from the Boltzmann statistics.  The $L^d$ factors arise from the choice of $i$ in (4.3).  The middle factors arise from (4.4) and the final factor accounts for the fact that any of the $n$ vertices in a polymer may be the first vertex $i$ in (4.3).

We write the subsum of (5.1) we've approximated and expressed by $e^\mu$ and in standard style approximate $\mu$ as follows (with $s(n)$ written as $s$)
\be
\mu = \sum_n \left[ sd\; \ell n(L) - (s\; \ell n(s) - s) - \left( \frac{sd}2 \right) \ell n(\beta n) - s\; \ell n(n) + s\; \ell n(2) \right].
\ee
We use a Lagrange multiplier $\al$ to conserve (5.2), and differentiate
\be	\frac d{ds} (\mu + \al sn) = 0.	\ee
Solving (5.5) and (5.2) together one gets
\be	\sum_n  \frac1{n^{d/2}} \ e^{\al n} = \frac 1 2 (\sqrt{\beta})^d	\ee
and
\be	s(n) = 2 \left( \frac L {\sqrt{\beta}}\right)^d   \frac 1{n^{(1+ d/2)} }\ e^{\al n} \ .	\ee 

We now restrict the above approximation to the trace on ${\cal H}_k$.  This involves considering sequences $r_i(n)$ satisfying
\be		\sum_n \ r_i(n)n = k 		\ee
with $r_i(n)$ satisfying
\be	r_i(n) \le s(n).  \ee
The $i$ labels such a sequence of $r_i(n)$.

Then
\be {\rm Tr} \left( e^{-\beta H} \right)_k \cong \sum_i \prod_{n=1} \left( \begin{array}{c}
s(n) \\
r_i(n)	\end{array} \right) \cdot e^\mu \ .
\ee
Basically we are selecting ways of choosing which cycles have spin up, and making sure for each such choice (5.8) holds, so there are total $k$ spins up.

We again approximate the sum in (5.10) by its biggest term, using a Lagrange multiplier to uphold (5.8).  We let $\tau$ be the Lagrange multiplier.  We get
\be	r(n) = s(n) \cdot \frac{e^{\tau n}}{1 + e^{\tau n}}	\ee
and
\be	\sum_n 2 \left( \frac L{\sqrt{\beta}}\right)^d \frac 1{n^{d/2}} \left( \frac{e^{\tau n}}{1 + e^{\tau n}} \right) = k \ . \ee

\bigskip
\bigskip
\section{The Picture}
\setcounter{equation}{0}

\ \ \ \ \ \ We consider three cases

\noindent
\underline{Case 1} \ \ \ \ \ $d=2$.

\noindent
\underline{Case 2}  \ \ \ \ \ $d=3$, and $\beta < < 1$.

\noindent
\underline{Case 3}  \ \ \ \ \ $d=3$, and $\beta > > 1$.

In cases 1 and 2 one has the equation (5.6) satisfied with $\al < 0$, $s(n)$ as given by (5.7).  Equations (5.6) and (5.7) are satisfied using only ``finite" $k$-cycles, there are no ``infinite" $k$-cycles.  There is not spontaneous magnetization.

In case 3, to satisfy (5.6) $\al$ must be greater than zero.  For given such $\beta$, as $L$ gets large one has the following limiting situation:
\be	\al = 0 + \ve	.	\ee
That is, $\al \rightarrow 0^+ $ as $L \rightarrow \infty$.
\be	s(n) = 2 \left( \frac L {\sqrt{\beta}}\right)^3   \frac 1{n^{5/2}}   \ee
for ``finite" $n$, and in addition a single ``infinite" $k$-cycle with $k$ given by
\be	k = N - 2N \frac 1{\beta^{3/2}} \sum \ \frac 1{n^{3/2}} \ .	\ee
This single ``infinite" $k$-cycle is needed to complement in the sum of (5.6) the contributions of ``finite" $k$-cycles.

It is perhaps easy for the reader to believe, and even deduce at the level of our calculations that the presence of the ``infinite" $k$-cycle yields spontaneous magnetization.  Alternatively we may use the arguement  of Section 2, picking $r$ satisfying (for $\beta >> 1$)
\be	r >> \frac 1{\beta^{3/2}} \ \sum \frac 1{n^{3/2}}	\ee

We have attempted some improvements to the estimates used, particularly in Section 5, but the ones we have considered have not changed the flow of the argument and results in any meaningful way.  We believe the picture we have presented herein is essentially correct, and that the key test and challenge to completing a rigorous presentation will be in proving a satisfactory form of (3.4), the central approximation.  The remaining steps, not unremniscent of a Peierls' argument,  may be easier to substantiate.

\bigskip
\noindent
\underline{Acknowledgement}.  I'd like to thank Joe Conlon for getting me started, Elliott Lieb for keeping me going, and Murph Goldberger for teaching me the ``What else could it be" type of argument.

\vfill\eject
\centerline{\underline{References}}

\begin{itemize}
\item[[1]]  P. Federbush, For the Quantum Heisenberg Ferromagnet, Some Conjectured Approximations, math-ph/0101017.
\item[[2]]  P. Federbush, For the Quantum Heisenberg Ferromagnet, A Polymer Expansion and its High T Convergence, math-ph/0108002.
\end{itemize}

\end{document}